\documentclass[12pt,reqno]{amsart}

\usepackage{fullpage}
\usepackage{graphicx}
\usepackage{amssymb, amsmath, amsxtra}

\newtheorem{thm}{Theorem}[section]
\theoremstyle{plain}
\newtheorem{cor}{Corollary}[section]
\theoremstyle{plain} 
\newtheorem{lem}{Lemma}[section]
\theoremstyle{plain}

\theoremstyle{plain}

\newtheorem{prop}{Proposition}[section]

\def\Rnum{{\mathbb R}}

\def\arctanh{{\rm arctanh}}

\def\Esp{{\mathcal{E}}}

\def\p{\partial}

\def\const{\text{const.}}

\def\tbox#1#2{$\genfrac{}{}{0pt}{}{\hbox{#1}}{\hbox{#2}}$}

\def\Ref#1{Ref.\cite{#1}}

\def\scrpt#1{$\scriptstyle {#1}$}

\begin{document}

\allowdisplaybreaks[3]

\title{
Conserved norms and related conservation laws\\ for multi-peakon equations
}

\author{
Elena Recio$^{1,2}$ \lowercase{\scshape{and}}
Stephen C. Anco$^1$
\\\\\lowercase{\scshape{
${}^1$Department of Mathematics and Statistics\\
Brock University\\
St. Catharines, ON L\scrpt2S\scrpt3A\scrpt1, Canada}} \\\\
\lowercase{\scshape{
${}^2$Department of Mathematics\\
Faculty of Sciences, University of C\'adiz\\
Puerto Real, C\'adiz, Spain, 11510}}\\
}

\begin{abstract}
All nonlinear dispersive wave equations in the general class 
$m_t+f(u,u_x)m+(g(u,u_x)m)_x =0$
are known to possess multi-peakon weak solutions. 
A classification is presented for families of multi-peakon equations in this class
that possess 
conserved momentum; 
conserved $H^1$ norm of $u$;
conserved $H^2$ norm of $u$;
conserved $L^2$ norm of $m$;
related conservation laws. 
The results yield, among others, two interesting wide families of equations:
$m_t + 2u_x h(u,u_x)m + u (h(u,u_x)m)_x=0$
for which the $H^1$ norm of $u$ is conserved; 
$m_t -\tfrac{1}{2}u_x h'(u)m + (h(u)m)_x =0$
for which the $L^2$ norm of $m$ is conserved. 
The overlap of these two families yields a singular equation 
which is nevertheless found to possess both smooth solitary wave solutions 
and peakon travelling wave solutions. 
\end{abstract}

\maketitle

\begin{center}
emails: 
elena.recio@uca.es, 
sanco@brocku.ca
\end{center}

\section{Introduction}\label{intro}

There has been a considerable amount of interest in the study of
wave-breaking equations belonging to the general class of 
nonlinear dispersive evolution equations given by 
\begin{equation}\label{fg-fam}
m_t + f(u,u_x)m + (g(u,u_x)m)_x =0, 
\quad
m=u-u_{xx} 
\end{equation}
for $u(t,x)$. 
This class includes all examples of multi-peakon equations known in the literature, particularly:
$b$-family equation ($f=(b-1)u_x$, $g=u$)
which contains both the Camassa--Holm and Degasperis--Procesi equations ($b=2,3$); 
Novikov equation ($f=uu_x$, $g=u^2$);
the modified Camassa--Holm equation ($f=0$, $g=u^2-u_x^2$); 
and their various nonlinear generalizations \cite{MiMu,GraHim,AncSilFre,HimMan2016a,HimMan2016b,AncRecGanBru,AncRec}. 
In addition to possessing multi-peakon solutions, 
these well-studied equations exhibit wave breaking 
in which certain smooth initial data yields solutions whose gradient $u_x$ 
blows up in a finite time 
\cite{Con-Esc-1998,Esc-Liu-Yin,Liu-Yin,Gui-Liu-Tian,Gui-Liu-Olv-Qu,Jia-Ni}. 

Recent work \cite{AncRec} has shown that every equation in the whole class \eqref{fg-fam} 
(for any smooth functions $f(u,u_x)$ and $g(u,u_x)$)
admits $N$-peakon weak solutions which consist of 
a linear superposition of peaked travelling waves 
\begin{equation}\label{multi-peakon}
u(t,x) = \sum_{i=1}^{N} a_i(t) \exp(-|x-x_i(t)|), 
\quad N = 1,2,\ldots
\end{equation}
having time-dependent amplitudes $a_i(t)$ and positions $x_i(t)$.
In particular, 
neither a Hamiltonian structure nor an integrability structure are necessary 
for the existence of $N$-peakon weak solutions for all $N\geq 1$. 

This interesting result raises the question of establishing 
a local well-posedness result, a global existence result for strong solutions, 
and a peakon stability result for weak solutions, 
which would apply to as wide of a subfamily of equations as possible 
in this class. 
For proving such analytical results, 
it will be essential to use conserved norms and other conservation laws
that are admitted by these equations. 

The present paper takes a first step by classifying 
families of nonlinear multi-peakon equations, 
contained in the general class \eqref{fg-fam} with $f,g\neq\const$, 
that possess:
(i) conserved momentum; 
(ii) conserved $H^1$ norm;
(iii) conserved $H^2$ norm;
(iv) related conservation laws. 
This classification reveals two interesting families of equations:
the first family is given by 
$f(u,u_x) = -\tfrac{1}{2}u_x h'(u)$, $g(u,u_x) = h(u)$, 
for which the $L^2$ norm of $m$ is conserved,
where $h(u)$ is an arbitrary smooth function of $u$;
the second family is given by 
$f(u,u_x) = u_x h(u,u_x)$, $g(u,u_x) = u h(u,u_x)$,
for which the $H^1$ norm of $u$ is conserved,
where $h(u,u_x)$ is an arbitrary smooth function of $u$ and $u_x$.

Several applications of the classification results are considered. 
One application is that conservation of the $L^2$ norm of $m$ for strong solutions 
is shown to yield an upper bound on the maximal amplitude and the maximal gradient of these solutions in terms of their initial data. 
Another application is that conservation of both momentum and the $H^1$ norm of $u$
is shown to imply the existence of a Hamiltonian structure 
and to rule out smooth solitary wave solutions. 
A third application is that conservation of both the $H^1$ norm of $u$ and the $L^2$ norm of $m$ 
leads to a multi-peakon equation that is shown to possess 
smooth solitary wave solutions as well as peakon solutions. 
This is the first such peakon equation ever identified. 

In section~\ref{conslaws}, 
we discuss how to obtain conservation laws for multi-peakon equations in the general class \eqref{fg-fam},
by using a modification the general multiplier method \cite{AncBlu2002a,AncBlu2002b,Anc-review}
combined with some tools from variational calculus \cite{Olv-book,Anc-review}.
This modified approach can be applied to other types of equations 
and should be of interest on its own 
because it yields only those conservation laws having a specified form for the conserved integral rather than all conservation laws with a general form for the multiplier. 
In section~\ref{classifications}, 
the classification results on $f(u,u_x)$ and $g(u,u_x)$ 
for each type of conservation law 
are derived, using some lemmas which are provided in an appendix. 
The three applications of these classification results are presented in section~\ref{applications}. 
Finally, 
some interesting examples of multi-peakon equations possessing a conserved momentum, 
a conserved $H^1$ norm, and a conserved $H^2$ norm
are summarized in section~\ref{concl},
where we also make some concluding remarks.

\section{Local conservation laws}\label{conslaws}

We will consider the general class of multi-peakon equations \eqref{fg-fam}
on an arbitrary spatial domain $\Omega\subseteq\Rnum$. 
In analysis, 
the domains most usually chosen are 
the whole line $\Omega=\Rnum$, 
with asymptotic decay conditions on $u(t,x)$ for $|x|\to\infty$; 
the half line $\Omega=\Rnum^+$, 
with either Dirichlet or Neumann boundary condition on $u(t,x)$ at $x=0$; 
a closed interval $\Omega =[0,1]$,
with periodic boundary conditions on $u(t,x)$ at $x=0,1$. 
The specific choice of domain will not play any role 
in finding local conservation laws. 

A local conservation law for a multi-peakon equation \eqref{fg-fam} 
is a continuity equation 
\begin{equation}\label{localCL}
D_t T + D_x\Phi = 0
\end{equation}
holding for all locally smooth solutions $u(t,x)$ of the multi-peakon equation,
with $x\in\Omega$ and $t\in [0,t_0)$ for some $t_0>0$,
where $T$ and $\Phi$ are differentiable functions that depend on 
$t$, $x$, $u$, and derivatives of $u$. 
Here $D$ denotes a total derivative (acting by the chain rule). 

Integration of a local conservation law over the spatial domain $\Omega$ yields
a global balance equation 
$\frac{d}{dt}\int_\Omega T\,dx = -\Phi\big|_{\p\Omega}$
which holds for all solutions $u(t,x)$ in a suitable function space $X$ on $\Omega$.
If the flux $\Phi\big|_{\p\Omega}$ vanishes 
after any asymptotic decay conditions or boundary conditions 
for solutions $u\in X$ are taken into account, 
then the balance equation yields a conserved integral 
\begin{equation}\label{globalCL}
\frac{d}{dt}\int_\Omega T\,dx = 0 . 
\end{equation}

A conserved integral \eqref{globalCL} is generally useful only 
if the conserved density $T(t,x,u,\p u,\p^2 u,\ldots)$ does not have the form 
$D_x\Theta(t,x,u,\p u,\p^2 u,\ldots)$ 
for some differentiable function $\Theta$, 
since otherwise $\int_\Omega T\, dx = \Theta\big|_{\p\Omega}$ 
reduces to flux terms that depend 
merely on the asymptotic decay or boundary conditions
posed for solutions $u\in X$. 
The global balance equation 
$\frac{d}{dt}\int_\Omega T\,dx = -\Phi\big|_{\p\Omega}$ 
thereby becomes an identity, with $\Phi= -D_t\Theta$. 
Consequently, 
a conservation law is said to be locally trivial \cite{Olv-bbok,Anc-review} when
$T(t,x,u,\p u,\p^2 u,\ldots) = D_x\Theta(t,x,u,\p u,\p^2 u,\ldots)$
and 
$\Phi(t,x,u,\p u,\p^2 u,\ldots) = -D_t\Theta(t,x,u,\p u,\p^2 u,\ldots)$
hold for all solutions $u\in X$.
Moreover, 
any non-trivial local conservation law can be changed 
an inessential way by the addition of 
a locally trivial density $D_x\Theta$ to $T$ and
a corresponding flux $-D_t\Theta$ to $\Phi$,
since $\int_\Omega T\,dx$ remains the same up to flux terms 
which will typically vanish due to the asymptotic decay or boundary conditions
posed for solutions $u\in X$. 
Accordingly, 
any two local conservation laws that differ by a locally trivial conservation law 
are said to be locally equivalent \cite{Olv-bbok,Anc-review}. 

The most relevant kind of non-trivial conservation laws for doing analysis
are conserved energies and conserved Sobolev norms. 
These belong to the class of local conservation laws \eqref{localCL} 
having the form 
\begin{equation}\label{CL:T}
T(u,u_x,m) \neq D_x\Theta(x,u,u_x)
\end{equation}
where, specifically,
\begin{align}
T & = u_x^2 + N(u)
\label{energyT}
\\
T & = u_{xx}^2 + N(u,u_x)
\label{gradenergyT}
\end{align}
comprise the general forms for a conserved energy density
and a conserved gradient energy density,
while, as particular cases, 
\begin{equation}\label{HnormT}
T = u_x^2 + \nu u^2, 
\quad 
T = u_{xx}^2 + \mu u_x^2 + \nu u^2,
\quad
\mu,\nu >0
\end{equation}
are the forms given by the densities 
for (weighted) $H^1$ and $H^2$ norms, respectively.
Another non-trivial conservation law of interest in the same class
is the momentum 
\begin{equation}\label{momentumT}
T = m = u-u_{xx}, 
\end{equation}
which has the locally equivalent form 
\begin{equation}
T = u . 
\end{equation}
A general method \cite{AncBlu2002a,AncBlu2002b,Anc-review}, 
using multipliers, 
is available to find all of these local conservation laws. 

We will use a modified version of the multiplier method, 
which can be applied, without loss of generality, 
because of the specific forms \eqref{energyT}--\eqref{momentumT} of $T$ 
characterizing energies and Sobolev norms as well as momentum. 
What is different about the modified method is that it will yield precisely those 
local conservation laws with a specified form for the conserved density, 
rather than all local conservation laws with a general form for the multiplier. 

It will be useful to write 
\begin{equation}\label{fg-eqn}
\Upsilon(u,u_x,m,m_x) = m_t+ f(u,u_x)m + D_x(g(u,u_x)m)
\end{equation}
for the expression defining a general multi-peakon equation \eqref{fg-fam}.
In particular, a locally smooth function $u(t,x)$ 
will belong to the space $\Esp$ of locally smooth solutions to this equation 
when and only when $\Upsilon=0$ for this function $u(t,x)$. 
Then any local conservation law admitted by a multi-peakon equation 
can be expressed as $(D_t T + D_x \Phi)|_\Esp=0$,
where a restriction to $\Esp$ is taken to mean that all $t$-derivatives of $m$ 
are eliminated through substitution of $m_t = -f(u,u_x)m -D_x(g(u,u_x)m)$. 
When $T$ has the form \eqref{CL:T}, 
we note that $(D_t T)|_\Esp$ will be a function of $u$, $u_x$, $m$, $m_x$, 
as well as $u_t$ and $u_{tx}$ (which will appear linearly). 
The relation $(D_x \Phi)|_\Esp= -(D_t T)|_\Esp$ then shows that 
$\Phi$ must be a function of at most $u$, $u_x$, $m$, $u_t$, and $u_{tx}$. 
Moreover, when $u(t,x)$ is an arbitrary locally smooth function, 
then $T(u,u_x,m)$ and $\Phi(u,u_x,m,u_t,u_{tx})$ will satisfy the identity 
\begin{equation}\label{chareqn}
D_t T +D_x \Phi = \Upsilon Q
\end{equation}
where $Q(u,u_x,m,u_t,u_{tx})$ is given by the relation 
\begin{equation}\label{multr}
Q = T_m -\Phi_{u_{tx}} .
\end{equation}
This identity \eqref{chareqn}--\eqref{multr} is called the characteristic equation for the conservation law,
and $Q$ is called the conservation law multiplier. 
The pair $(T,\Phi)$ is commonly called a conserved current. 

A one-to-one correspondence between multipliers $Q$ 
and locally equivalent conserved currents $(T+D_x\Theta,\Phi-D_t\Theta)$
can be established from the characteristic equation \eqref{chareqn}. 
This result follows from a general correspondence \cite{Olv-book,AncBlu2002b,Anc-review}
that holds between multipliers and locally equivalent conservation laws
for any partial differential equation having a generalized Cauchy--Kovalevskaya form. 
In the case of a multi-peakon equation $m_t + fm +(gm)_x=0$, 
this form is given by $u_{xxx} = u_x +(u_{txx} -u_t + (u_{xx}-u)(f +g_x))/g$.
However, because the leading derivative is an $x$-derivative of $u$, 
there will not necessarily be a one-to-one correspondence between 
multipliers $Q$ and locally equivalent conserved densities $T+D_x\Theta$. 

As an example, 
the multi-peakon equation \eqref{fg-fam} given by 
$f(u,u_x)=-2uu_x$, $g(u,u_x) = u^2-3u_x^2$
admits a purely spatial conservation law $(D_x\Phi)|_\Esp=0$
where the flux is $\Phi=(u^3-uu_x^2-g m+u_{tx})^2-(u^2u_x-u_x^3+u_t)^2$,
and the corresponding multiplier is $Q=2u(u_x^2-u^2)+2gm-2u_{tx}$.
Clearly, if this flux conservation law is added to any local conservation law, 
it will not change the conserved density but will change the multiplier. 

Nevertheless, 
the characteristic equation \eqref{chareqn}--\eqref{multr} can be used 
to determine when any Sobolev norm density \eqref{HnormT} is conserved, 
and also to find all conserved energy densities \eqref{energyT} and \eqref{gradenergyT}. 
The starting point is that $T(u,u_x,m)$ satisfies a local conservation law 
iff $D_t T - Q\Upsilon$ is a total $x$-derivative $D_x\Phi(u,u_x,m,u_t,u_{tx})$, 
for some multiplier $Q(u,u_x,m,u_t,u_{tx})$. 
From general results in variational calculus, 
we know that total $x$-derivatives $D_x\Phi(u,u_x,m,u_t,u_{tx})$ 
are characterized by \cite{Olv-book,Anc-review} 
belonging to the kernel of the spatial Euler operators
$\hat E_{u}$ and $\hat E_{u_t}$, which are defined by 
\begin{equation}\label{euler_op}
\hat E_v = \partial_{v} -D_x\partial_{v_x} + D_x^2 \partial_{v_{xx}} -D_x^3 \partial_{v_{xxx}} + \cdots
\end{equation}
for any variable $v$. 

\begin{lem}\label{TQ:conds}
A function $T(u,u_x,m)$ satisfies a local conservation law 
$(D_t T + D_x \Phi)|_\Esp=0$
for a multi-peakon equation $m_t+ f(u,u_x)m + D_x(g(u,u_x)m)=0$
iff 
\begin{align}
& \hat E_{u}(D_t T -Q\Upsilon) =0, 
\label{Eu:T}
\\
& \hat E_{u_t}(D_t T - Q\Upsilon) =0 
\label{Eut:T}
\end{align}
hold for some multiplier function $Q(u,u_x,m,u_t,u_{tx})$, 
when $u(t,x)$ is an arbitrary locally smooth function,
where $\Upsilon$ is expression \eqref{fg-eqn}. 
\end{lem}

These two conditions \eqref{Eu:T}--\eqref{Eut:T} 
will, in general, be polynomials in derivatives of $m$
and must hold for all locally smooth functions $u(t,x)$. 
Hence, 
the coefficients of each different monomial in derivatives of $m$ must vanish, 
which yields a linear overdetermined system of equations involving $T$, $Q$, $f$ and $g$. 
The example above shows that this system can have a non-trivial solution for $f$, $g$, and $Q$, when $T=0$. 
So, rather than try to find all solutions, 
we will take $Q$ to have the simplest possible form that allows 
the system to be solved with a given form for $T(u,u_x,m)$. 
We can do this by considering the terms 
$\hat E_{u}(D_t T)$ and $\hat E_{u_t}(D_t T)$
in the respective conditions \eqref{Eu:T}--\eqref{Eut:T},
and applying a derivative balancing and counting argument, 
as explained in the next section.

\section{Classification results}
\label{classifications} 

We will now proceed to determine families of nonlinear multi-peakon equations \eqref{fg-fam}
that admit local conservation laws in which $T$ has one of the forms \eqref{energyT}--\eqref{gradenergyT}, \eqref{HnormT}, \eqref{momentumT} 
that characterize energies, $H^1$ and $H^2$ Sobolev norms, and momentum, respectively. 

The computations will require solving Euler-operator equations of the form 
$\hat E_u(h_1 m + h_2 m^2 + h_3 m^3)=0$
and $\hat E_{u_t}(h_1 m + h_2 m^2 + h_3 m^3)=0$,
where $h_1$, $h_2$, $h_3$ are functions of $u$ and $u_x$. 
We give a derivation of the general solution, 
stated as Lemma~\ref{euler_eqn}, 
in the Appendix. 

We will only be interested in cases 
where at least one of the coefficient functions $f(u,u_x)$, $g(u,u_x)$ 
in the general multi-peakon equation \eqref{fg-fam} is non-constant. 
This condition, $f_u^2 + f_{u_x}^2 + g_u^2 + g_{u_x}^2 \neq 0$, 
is necessary and sufficient for a multi-peakon equation to be nonlinear. 

Classifications of conservation laws are typically carried out by taking into account the equivalence group of point transformations that preserve the general class of equations being studied. 
For the class of nonlinear multi-peakon equations \eqref{fg-fam}, 
the equivalence group will be small, 
because the relation $m=u-u_{xx}$ must be preserved.
In particular, 
a straightforward calculation using the Lie symmetry method \cite{2ndbook} 
shows that the equivalence group is given just by shifts in $f,g$. 

The results of the classifications can be viewed as a particular example of 
solving an inverse problem that consists of finding differential equations (within a certain class) 
that possess a conserved integral of a specified form. 
See Ref.\cite{PopBih} for some recent work on the general inverse problem for conservation laws.

\subsection{Conservation of momentum}

For $T=m$, we have simply $D_t T = m_t$,
and hence 
$\hat E_{u}(D_tT) = 0$, $\hat E_{u_t}(D_tT) = 1$. 
Then, from Lemma~\ref{TQ:conds}, 
the determining conditions \eqref{Eu:T}--\eqref{Eut:T} become
\begin{equation}\label{mom:Qdeteqns}
\hat E_{u}(m_tQ + (fm+D_x(gm))Q)=0, 
\quad
\hat E_{u_t}(m_tQ + (fm+D_x(gm))Q)=1 . 
\end{equation}
Now we look for the simplest form for $Q(u,u_x,m,u_t,u_{tx})$ 
that will allow these two equations to be solved. 
Since the term $m_tQ$ has one more $t$-derivative of $u$ in total 
compared to the term $(fm+D_x(gm))Q$, 
the simplest possible balance of $t$-derivatives on the left and right sides 
in each equation is obtained by putting 
\begin{equation}\label{mom:ut-balance}
\hat E_{u}(m_tQ)=0,
\quad
\hat E_{u_t}(m_tQ)=1 . 
\end{equation}
Then the simplest solution for this pair of equations 
can be obtained by taking $Q$ to have no dependence on $u_t$ and $u_{tx}$. 
Hence, we will put $Q=Q(u,u_x,m)$. 
Each of the equations \eqref{mom:ut-balance} then splits with respect to the variables
$m_x$, $m_{xx}$, $u_t$, $u_{tx}$, $m_t$, $m_{tx}$, $m_{txx}$, 
yielding a linear system of four equations.
The first three equations are found to reduce to 
$Q_u =0$, $Q_{u_x} =0$, $Q_m =0$, 
which implies $Q$ is a constant. 
Then the remaining equation in the split systems becomes
\begin{equation}
Q=1 .
\end{equation}
Substituting this expression for $Q$ back into the determining conditions \eqref{mom:Qdeteqns}, 
we get 
\begin{equation}\label{mom:euler_eqn}
\hat E_u(fm) =0 .
\end{equation}

By applying Lemma~\ref{euler_eqn} to the condition \eqref{mom:euler_eqn}, 
we obtain 
\begin{equation}\label{mom:f}
f(u,u_x) = u_x f_1(u^2-u_x^2) + \frac{f_0 u}{u^2-u_x^2}
\end{equation}
and
\begin{equation}\label{mom:Phi}
\Phi(u,u_x,m) = g(u,u_x)m +\tfrac{1}{2} F_1(u^2-u_x^2) + \tfrac{1}{2} f_0 \ln\Big(\frac{u-u_x}{u+u_x}\Big) + f_0 x
\end{equation}
where $f_1$ is an arbitrary function of $u^2-u_x^2$, 
$F_1$ is a function of $u^2-u_x^2$ such that $F_1'=f_1$, 
and $f_0$ is an arbitrary constant.

Therefore, 
the local conservation law for momentum is given by 
expressions \eqref{momentumT} and \eqref{mom:Phi} for the density and flux, 
respectively. 
This conservation law has the locally equivalent form 
\begin{equation}\label{mom:densflux}
T= u,
\quad
\Phi = g(u,u_x)m - u_{tx} +\tfrac{1}{2} F_1(u^2-u_x^2) + \tfrac{1}{2} f_0 \ln\Big(\frac{u-u_x}{u+u_x}\Big) + f_0 x .
\end{equation}

\begin{thm}\label{thm:mom}
Local conservation of momentum, $(D_t u + D_x\Phi)|_\Esp =0$, 
holds for the family of multi-peakon equations \eqref{fg-fam} 
with $f(u,u_x)$ having the form \eqref{mom:f}
and with $g(u,u_x)$ being arbitrary. 
The total momentum on a domain $\Omega$ 
\begin{equation}\label{mom_integral}
M[u] = \int_\Omega m\, dx 
\end{equation}
will be conserved $\frac{d}{dt}M = 0$ 
for all solutions $u\in X$ whose flux \eqref{mom:Phi} vanishes at $\p\Omega$. 
\end{thm}

\subsection{Conservation laws related to $H^1$ norm}

We start by considering $T=u_x^2 +u^2$, 
which is the density for the $H^1$ norm. 
Then we have 
$D_t T = 2 u_xu_{tx} +2uu_t$, 
and hence 
$\hat E_{u}(D_tT) = 2m_t$, $\hat E_{u_t}(D_tT) = 2m$. 
The determining conditions \eqref{Eu:T}--\eqref{Eut:T} 
from Lemma~\ref{TQ:conds} thereby become
\begin{equation}
\hat E_{u}(m_tQ + (fm+D_x(gm))Q)=2m_t, 
\quad
\hat E_{u_t}(m_tQ + (fm+D_x(gm))Q)=2m . 
\end{equation}
The simplest possible balance of $t$-derivatives on the left and right sides 
in each of these equation is obtained by 
taking $Q$ to have no dependence on $u_t$, $u_{tx}$, 
and putting 
\begin{equation}\label{H1:ut-balance}
\hat E_{u}(m_tQ)=2m_t,
\quad
\hat E_{u_t}(m_tQ)=2m . 
\end{equation}
Then, with $Q=Q(u,u_x,m)$, 
we find that the pair of equations \eqref{H1:ut-balance} 
splits with respect to the variables
$m_x$, $m_{xx}$, $u_t$, $u_{tx}$, $m_t$, $m_{tx}$, $m_{txx}$, 
yielding a linear system of four equations.
From the first three equations, we obtain 
$Q_u =2$, $Q_{u_x} =0$, $Q_m =0$, 
and then the remaining equation in the split system yields
\begin{equation}
Q=2u . 
\end{equation}
We now substitute $Q$ back into the determining conditions \eqref{mom:Qdeteqns},
yielding 
$\hat E_u((fm+D_x(gm))u) =0$. 
This equation can be simplified through integration by parts 
on the term $uD_x(gm)= D_x(ugm)-u_xgm$,
which gives
\begin{equation}\label{H1:euler_eqn}
\hat E_u((uf-u_xg)m) =0 . 
\end{equation}

By applying Lemma~\ref{euler_eqn} to the condition \eqref{H1:euler_eqn}, 
we obtain 
\begin{equation}\label{H1:fgrel}
uf(u,u_x)-u_xg(u,u_x) = u_x h_1(u^2-u_x^2) + h_0\frac{u}{u^2-u_x^2}
\end{equation}
and
\begin{equation}\label{H1:Phi}
\Phi(u,u_x,u_{tx},m) -2u g(u,u_x)m + 2uu_{tx} 
= H_1(u^2-u_x^2) + h_0 \ln\Big(\frac{u-u_x}{u+u_x}\Big) + 2h_0 x
\end{equation}
where $h_1=H_1'$ is an arbitrary function of $u^2-u_x^2$, 
and $h_0$ is an arbitrary constant.
From the relation \eqref{H1:fgrel}, we get 
$f = u_x h + u_x \tilde{h}$ and $g = u h -u \tilde{h}$
where $h$ is an arbitrary function of $u$ and $u_x$,
and where $\tilde{h}$ is given by
\begin{equation}
\tilde{h}(u,u_x) = \frac{\tfrac{1}{2} h_1(u^2-u_x^2)}{u} + \frac{\tfrac{1}{2}h_0}{u_x(u^2-u_x^2)} . 
\end{equation}
This yields
\begin{align}
f(u,u_x) & = u_x h(u,u_x) + \frac{\tfrac{1}{2}u_x h_1(u^2-u_x^2)}{u} + \frac{\tfrac{1}{2}h_0}{u^2-u_x^2} , 
\label{H1:f}
\\
g(u,u_x) & = u h(u,u_x) -\tfrac{1}{2}h_1(u^2-u_x^2) -\frac{\tfrac{1}{2}h_0 u}{u_x(u^2-u_x^2)} .
\label{H1:g}
\end{align}
Therefore, 
the local conservation law for the $H^1$ norm (squared) is given by 
the density and flux expressions 
\begin{align}
&\begin{aligned}
T= u_x^2 + u^2,
\end{aligned}
\label{H1:dens}
\\
&\begin{aligned}
\Phi & = -2uu_{tx} + 2u^2 h(u,u_x) m -uh_1(u^2-u_x^2)m - h_0\frac{u^2m}{u_x(u^2-u_x^2)} +H_1(u^2-u_x^2) 
\\&\qquad
+ h_0 \ln\Big(\frac{u-u_x}{u+u_x}\Big) + 2h_0 x .
\end{aligned}
\label{H1:flux}
\end{align}

\begin{thm}\label{thm:H1}
Local conservation of the $H^1$ density,
$(D_t(u_x^2+u^2) + D_x\Phi)|_\Esp =0$,
holds for the family of multi-peakon equations \eqref{fg-fam} 
with $f(u,u_x)$ and $g(u,u_x)$ given by the respective forms \eqref{H1:f}--\eqref{H1:g}. 
The $H^1$ norm on a domain $\Omega$ 
\begin{equation}\label{H1_integral}
||u||_{H^1} = \Big( \int_\Omega u_x^2 + u^2 \, dx \Big)^{1/2}
\end{equation}
will be conserved $\frac{d}{dt}||u||_{H^1} = 0$ 
for all solutions $u\in X$ whose flux \eqref{H1:flux} vanishes at $\p\Omega$. 
\end{thm}

We will next consider related conservation laws for energy having the form \eqref{energyT} 
which includes a weighted $H^1$ norm when $N(u)=\nu u^2$ with $\nu\neq 1$. 
By repeating the previous steps, we have 
\begin{equation}\label{energy:ut-balance}
\hat E_{u}(m_tQ)=2m_t +(N''(u)-2)u_t,
\quad
\hat E_{u_t}(m_tQ)=2m + N'(u)-2u
\end{equation}
with $Q=Q(u,u_x,m)$. 
The solution of this pair of equations can be shown to be given by 
$Q=2u$ and $N=u^2$. 
As a result, 
the energy density $T=u_x^2+N(u)=u_x^2+u^2$ reduces to the $H^1$ norm density,
and we do not obtain a new local conservation. 

Finally, 
it is interesting to look at the overlap of the conditions under which 
the local conservation laws for both the momentum and the $H^1$ density 
hold. 
The following result is obtained by equating the forms for $(f(u,u_x),g(u,u_x))$ 
in Theorems~\ref{thm:mom} and~\ref{thm:H1}. 

\begin{cor}\label{thm:mom-H1}
Local conservation laws for both momentum and the $H^1$ norm 
hold for the family of multi-peakon equations \eqref{fg-fam} 
given by 
\begin{align}
f(u,u_x) & = u_x f_1(u^2-u_x^2) + \frac{f_0 u}{u^2-u_x^2} , 
\label{momH1:f}
\\
g(u,u_x) & = uf_1(u^2-u_x^2) +h_1(u^2-u_x^2) + \frac{f_0 u^2+h_0 u}{u_x(u^2-u_x^2)} , 
\label{momH1:g}
\end{align}
where $f_1,h_1$ are arbitrary functions of $u^2-u_x^2$,
and $f_0,h_0$ are arbitrary constants. 
\end{cor}

\subsection{Conservation laws related to $H^2$ norm}

We start by considering the density for the $H^2$ norm, 
$T=u_{xx}^2 +u_x^2 +u^2$. 
This gives $D_tT = -2u_{xx}m_t +2uu_t +D_x(2u_xu_t)$,
and hence we have 
$\hat E_{u}(D_tT) = 2(u_t-m_{txx})$, $\hat E_{u_t}(D_tT) = 2(u-m_{xx})$. 
The determining conditions \eqref{Eu:T}--\eqref{Eut:T} 
from Lemma~\ref{TQ:conds} are then given by 
\begin{align}
& \hat E_{u}(m_tQ + (fm+D_x(gm))Q)=2(u_t-m_{txx}), 
\label{H2:eq1}
\\
& \hat E_{u_t}(m_tQ + (fm+D_x(gm))Q)= 2(u-m_{xx}) . 
\label{H2:eq2}
\end{align}
Because the right side of equation \eqref{H2:eq1} is linear in first-order $t$-derivatives of $u$, 
the simplest possible balance of $t$-derivatives in this equation 
is obtained if $Q$ has no dependence on $u_t$ and $u_{tx}$. 
Similarly, because the right side of equation \eqref{H2:eq2} contains no $t$-derivatives of $u$, 
the simplest possible balance of $t$-derivatives is again obtained 
if $Q$ has no dependence on $u_t$ and $u_{tx}$. 
Hence, we will take $Q=Q(u,u_x,m)$. 
This possible balance of $t$-derivatives on both sides of each equation 
requires putting 
\begin{equation}\label{H2:ut-balance}
\hat E_{u}(m_tQ)=2(u_t-m_{txx}), 
\quad
\hat E_{u_t}(m_tQ)=2(u-m_{xx}) . 
\end{equation}
These two equations \eqref{H2:ut-balance} 
then split with respect to the variables
$m_x$, $m_{xx}$, $u_t$, $u_{tx}$, $m_t$, $m_{tx}$, $m_{txx}$, 
yielding a linear system of five equations.
When this system is simplified, it becomes
\begin{equation}
Q_u =0,
\quad
Q_{u_x} =0,
\quad
Q_m =2,
\quad
Q=2(m-u) . 
\end{equation}
which has no solution. 
Hence, the determining conditions \eqref{H2:eq1}--\eqref{H2:eq2}
have no solution when $Q$ is a function of $u$, $u_x$, $m$.

The next simplest possibility for balancing $t$-derivatives 
in the determining conditions \eqref{H2:eq1}--\eqref{H2:eq2}
would be to try $Q=Q_0(u,u_x,m) + Q_1(u,u_x,m)u_{tx}$
which is linear in $u_{tx}$, 
as motivated by the example presented in section~\ref{conslaws}. 
Repeating the previous steps, 
we reach the same conclusion that the determining conditions \eqref{H2:eq1}--\eqref{H2:eq2}
have no solution when $Q$ is a function of $u$, $u_x$, $m$, and linear in $u_{tx}$. 

This shows that local conservation of the $H^2$ density 
cannot hold with a multiplier $Q=Q_0(u,u_x,m) + Q_1(u,u_x,m)u_{tx}$
for any multi-peakon equation \eqref{fg-fam}. 
The question of whether a multiplier with the general form $Q(u,u_x,m,u_t,u_{tx})$
could work will be left to a subsequent paper. 

We will now turn to investigate more general conservation laws 
given by the form \eqref{gradenergyT} for $T$ that characterizes a gradient energy density. 
This form includes a weighted $H^2$ density when $N(u,u_x)=bu_x^2+au^2$ with $b\neq 1$ or $a\neq 1$. 

For $T=u_{xx}^2 +N(u,u_x)$, 
we have $D_tT = 2u_{xx}(u_t-m_t) +u_t N_{u} +u_{tx} N_{u_x}$,
which yields
\begin{equation}
\begin{aligned}
\hat E_{u}(D_tT) & = 2(u_t-m_t-m_{txx}) +u_t(N_{uu} -D_x N_{uu_x})-u_{tx}D_x N_{u_xu_x} +(m_t-u_t)N_{u_xu_x} 
\\
&= A_1(u,u_x,m,u_t,u_{tx},m_t,m_{tx})
\end{aligned}
\end{equation}
and 
\begin{equation}
\hat E_{u_t}(D_tT) = 2(u-m-m_{xx}) +N_{u} -D_x N_{u_x} = A_2(u,u_x,m,m_{xx}) .
\end{equation}
Hence the determining conditions \eqref{Eu:T}--\eqref{Eut:T} 
from Lemma~\ref{TQ:conds} are given by 
\begin{equation}\label{gradenergy:eqs}
\hat E_{u}(m_tQ + (fm+D_x(gm))Q)= A_1, 
\quad
\hat E_{u_t}(m_tQ + (fm+D_x(gm))Q)= A_2. 
\end{equation}
Using the same argument considered in the $H^2$ density case, 
we will take $Q=Q(u,u_x,m)$ and put 
\begin{equation}\label{gradenergy:ut-balance}
\hat E_{u}(m_tQ) =A_1,
\quad
\hat E_{u_t}(m_tQ)=A_2 . 
\end{equation}
These two equations \eqref{gradenergy:ut-balance} 
split with respect to the variables
$m_x$, $m_{xx}$, $u_t$, $u_{tx}$, $m_t$, $m_{tx}$, $m_{txx}$, 
yielding a linear system which, 
after simplification, is given by seven equations
\begin{gather}
Q_{u_x} =0,
\quad
Q_m -2 =0,
\\
N_{u_x u_x u_x}=0,
\quad
N_{u_x u_x u} =0,
\quad
N_{u u}-u_x N_{u_x u u} -N_{u_x u_x} +2=0,
\\
Q_u -N_{u_x u_x} +4=0,
\quad
Q+u_x N_{u u_x}-N_{u}+2(u-m)=0 .
\end{gather}
This system is consistent and can be straightforwardly integrated:
\begin{align}
& N = \mu u_x^2 +(\mu -1)u^2 +2 \nu u+ D_x\Theta(u),
\label{gradenergy:N}
\\
& Q=2(m+(\mu -2)u+ \nu ) .
\label{gradenergy:Q}
\end{align}
We now substitute $Q$ back into the determining conditions \eqref{gradenergy:eqs},
which reduce to the equation 
$\hat E_u((fm+D_x(gm))(m+(\mu -2)u+ \nu )) =0$. 
This equation can be simplified by expanding and integrating by parts 
on the term 
$(m+(\mu -2)u)D_x(gm) = \tfrac{1}{2}m^2D_x g -(\mu -2)u_x gm +D_x(((\mu -2)u + \tfrac{1}{2}m)gm)$, 
so then we have
\begin{equation}\label{gradE:euler_eqn}
\hat E_u(((\mu -2)(uf-u_xg) +\nu f)m +(f+\tfrac{1}{2}(D_x g))m^2) =0 . 
\end{equation}

Applying Lemma~\ref{euler_eqn} to condition \eqref{gradE:euler_eqn}, 
we obtain three different (non-overlapping) solutions: 
\begin{align}
&
\nu =0,
\quad
\mu =2,
\quad
f=-\tfrac{1}{2}u_x h'(u),
\quad
g=h(u) ; 
\label{gradenergy3}
\\
&
\nu \neq 0,
\quad
\mu =2,
\quad
f= \alpha u_x,
\quad
g = -2\alpha u+\beta ;
\label{gradenergy2}
\\
& 
\nu = (\mu -2)\beta,
\quad
\mu\neq 2,
\quad
f=\alpha u_x/(u+\beta)^3,
\quad
g=\alpha/(u+\beta)^2 .
\label{gradenergy1}
\end{align}

Then from expressions \eqref{gradenergy:N}--\eqref{gradenergy:Q}, 
we find that the density and the flux in the resulting the local conservation law for the gradient energy are given by 
\begin{align}
&\begin{aligned}
T= u_{xx}^2 +\mu  u_x^2 + (\mu -1)u^2+2\nu u, 
\label{gradenergy:dens}
\end{aligned}
\\
&\begin{aligned}
\Phi & = 2((1-\mu)u-\nu)u_{tx} -2u_x u_t +(2((\mu-2)u+\nu)+m)m g
\\&\qquad
+((2-\mu)(u^2-u_x^2) -\nu u) g +\tfrac{1}{2}((\mu-2)u +\nu)u_x^2 g_u +\nu G,
\end{aligned}
\label{gradenergy:flux}
\end{align}
with $G=\int g\, du$. 

\begin{thm}\label{thm:gradE}
Local conservation of the gradient energy density,
$(D_t(u_{xx}^2 +\mu  u_x^2 + (\mu -1)u^2+2\nu u) + D_x\Phi)|_\Esp =0$,
holds for three families of multi-peakon equations \eqref{fg-fam},
where $f(u,u_x)$ and $g(u,u_x)$ have the respective forms 
\eqref{gradenergy3}, \eqref{gradenergy2}, \eqref{gradenergy1}.  
The gradient energy on a domain $\Omega$ 
\begin{equation}\label{gradE_integral}
E[u] = \int_\Omega u_{xx}^2 +\mu  u_x^2 + (\mu -1)u^2+2\nu u \, dx
\end{equation}
will be conserved $\frac{d}{dt}E = 0$ 
for all solutions $u\in X$ whose flux \eqref{gradenergy:flux} vanishes at $\p\Omega$. 
\end{thm}

There are two important special cases of the gradient energy \eqref{gradE_integral}. 

The first case is $\nu=0$, $\mu\neq 2$, 
where the gradient energy density reduces to a weighted $H^2$ density 
\begin{equation}\label{wH2:T}
T=u_{xx}^2 +\mu  u_x^2 + (\mu -1)u^2
\end{equation}
with the constant $\mu$ controlling the weights of $u_x^2$ and $u^2$. 
This special case occurs for the third family \eqref{gradenergy1}
when $\beta=0$. 
The flux in this case is given by 
\begin{equation}\label{wH2:Phi}
\Phi = 2(1-\mu)u u_{tx} -2u_x u_t +\alpha(\mu-1+m/u)^2 . 
\end{equation}
Then resulting local conservation law yields
\begin{equation}\label{wH2_integral}
\frac{d}{dt}||u||^2_{H^2,\mu}= -\Phi\big|_{\p\Omega}
\end{equation}
showing that the weighted $H^2$ norm 
\begin{equation}\label{H2_integral}
||u||_{H^2,\mu} = \Big( \int_\Omega u_{xx}^2 +\mu  u_x^2 + (\mu -1)u^2 \, dx \Big)^{1/2}
\end{equation}
will be conserved on a domain $\Omega$
when the flux \eqref{wH2:Phi} of $u\in X$ vanishes at $\p\Omega$. 

The second case is a specialization of the weighted $H^2$ norm \eqref{H2_integral}
given by $\mu=2$,
which occurs for the first family \eqref{gradenergy3}. 
In this case, 
the gradient energy conservation law has the locally equivalent form 
\begin{equation}\label{L2}
T = m^2 = (u-u_{xx})^2,
\quad
\Phi  = g(u) m^2 . 
\end{equation}
This local conservation law \eqref{L2} yields
\begin{equation}\label{L2_integral}
\frac{d}{dt}||m||^2_{L^2}= -\Phi\big|_{\p\Omega}
\end{equation}
which represents conservation of the $L^2$ norm of $m$ on a domain $\Omega$
when the flux $\Phi=g(u)m^2$ of $u\in X$ vanishes at $\p\Omega$. 

Finally, 
we consider the overlap of the conditions under which 
the local conservation law for the gradient energy holds
along with the local conservation laws for the $H^1$ norm and the momentum. 
By equating the form for $(f(u,u_x),g(u,u_x))$ in Theorem~\ref{thm:gradE} 
with the forms for $(f(u,u_x),g(u,u_x))$ in Theorems~\ref{thm:mom} and~\ref{thm:H1}, 
we obtain the following result. 

\begin{cor}\label{thm:gradE-H1-mom}
(i) None of the multi-peakon equations in the family \eqref{fg-fam} 
possess local conservation laws for both momentum and gradient energy. 
(ii) Local conservation laws for gradient energy and the $H^1$ norm 
hold for the family of multi-peakon equations \eqref{fg-fam} 
given by 
\begin{equation}\label{gradEH1:fg}
f(u,u_x) =\alpha u_x/u^3,
\quad
g(u,u_x) =\alpha/u^2
\end{equation}
where $\alpha,\beta$ are arbitrary constants. 
The gradient energy density and flux are respectively given by expressions \eqref{wH2:T} and \eqref{wH2:Phi}, where $\mu $ is an arbitrary constant. 
In the case $\mu =2$, this conservation law has the locally equivalent form \eqref{L2}
which is a local conservation law for the $L^2$ norm of $m$. 
\end{cor}

\section{Applications}\label{applications}

Three applications using the classification results stated In 
Theorems~\ref{thm:mom}, ~\ref{thm:H1}, and~\ref{thm:gradE}
will now be considered. 
We will work with 
smooth solutions $u(t,x)$ of the multi-peakon equations \eqref{fg-fam}
on the real line. 
All of the results carry over to strong solutions 
$u(t,x)\in C^0([0,\tau);H^s(\Rnum)) \cap C^1([0,\tau);H^{s-2}(\Rnum))$ 
with $s>\tfrac{7}{2}$,
for which Sobolev embedding ensures that $u$, $u_x$, $m$, $m_x$, and $u_t$ 
are continuous functions.

\subsection{Maximal amplitude and gradient of solutions}

Consider the family \eqref{gradenergy3} of multi-peakon equations \eqref{fg-fam}, 
in which the $L^2$ norm of $m$ is conserved. 
We will now show that this norm controls both 
the maximal amplitude and the maximal gradient of solutions $u(t,x)$. 

We start from an inequality obtained in \Ref{ConEsc}: 
\begin{equation}\label{u2-ineqn}
2u(t,x)^2 = 2\left( \int_{-\infty}^x uu_x\,dx - \int_x^{\infty} uu_x\,dx \right)
< \int_{-\infty}^x (u^2+u^2_x)\,dx + \int_x^{\infty} (u^2+u_x^2)\,dx 
= ||u||^2_{H^1}
\end{equation}
where the strict inequality comes from the fact that $u^2=u_x^2$ cannot hold 
for smooth functions with finite $H^1$ norm on the real line. 
By repeating this argument for the $x$-derivative of $u(t,x)$, we likewise have 
\begin{equation}\label{ux2-ineqn}
2u_x(t,x)^2 < ||u_x||^2_{H^1} .
\end{equation}
Note that we can choose different points $x\in\Rnum$ in each of these inequalities \eqref{u2-ineqn} and \eqref{ux2-ineqn}. 
This yields 
\begin{equation}
2(u(t,x_1)^2+ u_x(t,x_2)^2) < ||u||^2_{H^1} + ||u_x||^2_{H^1} 
= \int_{-\infty}^{\infty} (u^2+2u_x^2+u_{xx}^2 )\,dx 
= \int_{-\infty}^{\infty} m^2\,dx 
= ||m||^2_{L^2}
\end{equation}
since $2u_x^2 = 2(uu_x)_x - 2uu_{xx}$. 

Now, because the $L^2$ norm of $m$ is conserved, 
we conclude that 
\begin{equation}
\sup_{x\in\Rnum} |u(t,x)| < \tfrac{1}{\sqrt{2}} ||u_0||_{H^1} <  ||m_0||_{L^2},
\quad
\sup_{x\in\Rnum} |u_x(t,x)| < ||m_0||_{L^2} 
\end{equation}
with $m_0(x)=u_0(x)-u_0''(x)$,
where $u_0(x)=u(0,x)$ is the initial data for the Cauchy problem of 
the family \eqref{gradenergy3} of multi-peakon equations \eqref{fg-fam}.
These two inequalities give a priori bounds on the amplitude and gradient of the solutions.

\subsection{Hamiltonian structure and solitary waves}

Both the momentum and the $H^1$ norm are conserved in the family \eqref{momH1:f}--\eqref{momH1:g} of multi-peakon equations \eqref{fg-fam}. 
As a first observation, this family can be seen to coincide with 
the Hamiltonian family of multi-peakon equations derived in \Ref{AncRec}. 
Particular equations in the family include the Camassa--Holm equation 
and the modified Camassa--Holm equation. 

In the case of the Camassa--Holm equation, 
it is known \cite{Len} that all solitary travelling waves are given by peakons, 
namely, no smooth solitary waves exist. 
We will now show that the same result extends to all of the non-singular equations 
\begin{equation}\label{fg_CH_mCH_fam}
m_t +u_x f_1(u^2-u_x^2)  m + ((u f_1(u^2-u_x^2)  + g_1(u^2-u_x^2))m)_x=0 
\end{equation}
in the Hamiltonian family \eqref{momH1:f}--\eqref{momH1:g},
where $f_1$ and $g_1$ are smooth functions of $u^2 -u_x^2$. 

Consider a general travelling wave $u=U(\xi)$ with $\xi=x-ct$, 
where $c=\const\neq0$ is the wave speed. 
The Hamiltonian peakon equations \eqref{fg_CH_mCH_fam} then reduce to 
a third-order nonlinear ODE 
\begin{equation}
-c(U'-U''') + U'(U-U'')f_1(U^2-U'{}^2) + ((U-U'')(Uf_1(U^2-U'{}^2) + g_1(U^2-U'{}^2)))' 
=0 . 
\end{equation}
We can obtain two first integrals for this ODE 
by using the conservation laws for the momentum and the $H^1$-norm of solutions
in the following way \cite{PrzAnc}. 

\begin{lem}\label{lem:FI}
If the conserved density $T$ and the spatial flux $\Phi$ 
in a local conservation law $D_t T + D_x\Phi=0$ 
do not contain $t$ and $x$ explicitly, 
then for travelling waves $u=U(\xi)$ we see that 
$D_t= -c\frac{d}{d\xi}$ and $D_x= \frac{d}{d\xi}$
yields
$\frac{d}{d\xi}(\Phi- c T)=0$. 
This is a first integral, which gives $\Phi- c T=0$. 
It has the physical meaning of the spatial flux 
in a reference frame moving with speed $c$
(namely, the rest frame of the travelling wave). 
\end{lem}

Thus, from the momentum density and flux \eqref{mom:densflux},
we obtain the first integral 
\begin{equation}\label{FI_mom}
-c(U-U'') + \tfrac{1}{2} F_1+ (U-U'')(Uf_1 + g_1) =c_1=\const 
\end{equation} 
where
\begin{equation}
F_1=\int_{0}^{U^2-U'{}^2} f_1(y)\,dy 
=(U^2-U'{}^2)\tilde F_1(U^2-U'{}^2), 
\quad
\tilde F_1 = \int_0^1 f_1((U^2-U'{}^2)\lambda)\,d\lambda . 
\end{equation} 
From the $H^1$ density and flux \eqref{H1:dens}--\eqref{H1:flux}, 
we similarly obtain another first integral 
\begin{equation}\label{FI_H1}
-c(U^2+U'{}^2) + 2cUU'' - G_1 + 2U(U-U'')(Uf_1 + g_1) =c_2=\const
\end{equation} 
where
\begin{equation}
G_1=\int_{0}^{U^2-U'{}^2} g_1(y)\,dy 
=(U^2-U'{}^2)\tilde G_1(U^2-U'{}^2), 
\quad
\tilde G_1 = \int_0^1 g_1((U^2-U'{}^2)\lambda)\,d\lambda . 
\end{equation} 
We can now combine these two first integrals to get a first-order nonlinear ODE
\begin{equation}\label{travelling_wave_ODE}
(U'{}^2-U^2)(U \tilde F_1 +\tilde G_1-c) = 2c_1 U -c_2 .
\end{equation}
The smooth solutions $U(\xi)$ of this ODE 
are all smooth travelling waves of the Hamiltonian peakon equations \eqref{fg_CH_mCH_fam}. 

Solitary travelling waves are singled out by imposing the asymptotic decay condition
that $U$ and its derivatives go to zero for large $|\xi|$. 
Applying this condition to both first integrals \eqref{FI_mom} and \eqref{FI_H1}, 
and using the property that $\tilde G_1(0)=g_1(0)$ and $\tilde F_1(0)=f_1(0)$ are non-singular, 
we find $c_1=c_2=0$. 
Then the travelling-wave ODE \eqref{travelling_wave_ODE} becomes 
$(U'{}^2-U^2)(U \tilde F_1 +\tilde G_1-c) = 0$, 
which thereby implies $U'{}^2=U^2$ or $U \tilde F_1 +\tilde G_1 = c$. 
The first possibility directly leads to $U=0$ 
when asymptotic decay is imposed along with smoothness of $U(\xi)$;
the second possibility will be consistent with asymptotic decay 
iff $g_1(0)=c\neq 0$, 
but then wave speed will be fixed rather than arbitrary. 
This establishes the following non-existence result. 

\begin{prop}\label{prop:fg-fam-travellingwaves}
For the Hamiltonian family \eqref{fg_CH_mCH_fam} of multi-peakon equations, 
no smooth solitary travelling wave solutions exist either with arbitrary wave speed, 
or with a fixed wave speed if $g_1(0)=0$. 
\end{prop}

\subsection{Solitary waves and peakons}
For the one-parameter family \eqref{gradEH1:fg} of multi-peakon equations \eqref{fg-fam}, 
both the $H^1$ norm of $u$ and the $L^2$ norm of $m$ are conserved. 
This family can be written in the form 
\begin{equation}\label{fg_singular_fam}
m_t + u_x u^{-3} m +  (u^{-2} m)_x=0 
\end{equation}
after a scaling of $t$. 

The multi-peakon equation \eqref{fg_singular_fam}
has a singularity at $u=0$. 
Nevertheless, we will show that it possesses a smooth solitary wave solution,
as well as a peakon travelling wave solution. 
The effect of the singularity is that 
the amplitude of these solutions will be seen to diverge as the wave speed goes to zero. 

We will start by considering smooth travelling waves $u=U(\xi)$ with $\xi=x-ct$, 
where $c=\const\neq0$ is the wave speed. 
The peakon equation \eqref{fg_singular_fam} then reduces to 
a third-order nonlinear ODE 
\begin{equation}\label{travelling_wave_ODE2}
-c(U'-U''') + U'(U-U'')U^{-3} + ((U-U'')U^{-2})' 
=0 . 
\end{equation}
Two first integrals for this ODE 
arise by applying Lemma~\ref{lem:FI} to the conservation laws for 
the $H^1$-norm of $u$ and the $L^2$ norm of $m=u-u_{xx}$. 

The density and flux \eqref{L2} of the $L^2$ norm 
yield the first integral 
\begin{equation}\label{FI2_L2}
(U-U'')^2(c-1/U^2) =c_1=\const
\end{equation} 
while the $H^1$ density and flux \eqref{H1:dens}--\eqref{H1:flux} 
yield another first integral 
\begin{equation}\label{FI2_H1}
-c(U^2+U'{}^2) + 2cUU'' + 2(U-U'')/U =c_2=\const . 
\end{equation} 
By combining these two first integrals, we get a first-order nonlinear ODE
\begin{equation}\label{solitary_wave_ODE}
(c(U^2 -U'{}^2) -c_2)^2 = 4c_1(1-c U^2). 
\end{equation}
The smooth solutions $U(\xi)$ of this ODE 
are all smooth travelling waves of peakon equation \eqref{fg_singular_fam}. 

To obtain the solitary wave solutions,  
we first impose the asymptotic decay condition that 
$U$ and its derivatives go to zero for large $|\xi|$ 
in the ODE \eqref{solitary_wave_ODE}. 
This gives $c_1= \tfrac{1}{4}c_2^2$,
and hence the ODE becomes 
$U'{}^2 = U^2 +(c_2/c)\big(\sqrt{1-c U^2} -1\big)$, 
where we have chosen the sign of the square root so that $U'=0$ when $U=0$. 
We next require that the quadrature of this ODE gives 
\begin{equation}\label{solitary_wave_tail}
U\rightarrow 0 
\quad\text{as}\quad |\xi|\rightarrow \infty
\end{equation}
so that the solutions $U(\xi)$ exhibit an asymptotically decaying tail. 
To impose this requirement, 
we rewrite the right-hand side of the ODE by expressing 
$\sqrt{1-c U^2} -1 = -cU^2/(\sqrt{1-c U^2} +1)$,
whereby 
\begin{equation}\label{solitary_wave_ODE2}
U'{}^2 = U^2\frac{1-c_2+\sqrt{1-c U^2}}{1+\sqrt{1-c U^2}} . 
\end{equation}
Then for $U\sim 0$ we see that $U'{}^2 \sim (1-c_2/2)U^2$,
whose quadrature is $\pm\xi \sim \frac{\sqrt{2}}{\sqrt{2-c_2}}\ln(U)$,
yielding $|\xi|\rightarrow \infty$ as $U\rightarrow 0$ if $2-c_2>0$. 
Hence, we will put $c_2 = 2-b^2$, with $b\neq 0$. 
The ODE is now given by 
\begin{equation}\label{solitary_wave_ODE3}
U'{}^2 = U^2\frac{b^2-1+\sqrt{1-c U^2}}{1+\sqrt{1-c U^2}} . 
\end{equation}
Last, we want the solutions $U(\xi)$ to possess a peak at some point $\xi=\xi_0$. 
This requires that $U' =0$ holds for some $U\neq 0$. 
The ODE yields the condition $\sqrt{1-c U^2}=1-b^2$,
which can hold if and only if $0<|b|<1$ and $c>0$. 
Then the peak will be given by 
\begin{equation}\label{solitary_wave_peak}
|U|_{\max} = \frac{b\sqrt{2-b^2}}{\sqrt{c}},
\quad
0<b<1,
\quad
c>0
\end{equation}
which satisfies $0<\sqrt{c}|U|_{\max}<1$. 

It is straightforward to show that all solutions of the ODE \eqref{solitary_wave_ODE3}
with $c>0$ and $0<b<1$ describe solitary waves 
having a peak \eqref{solitary_wave_peak} 
and an asymptotic decaying tail \eqref{solitary_wave_tail}.
In particular, this ODE can be explicitly integrated, yielding 
\begin{equation}\label{solitarywave1}
\pm (\xi-\xi_0) = 
\arctanh\bigg(\frac{\sqrt{2B} A}{A+B}\bigg)
- (1/(\sqrt{2}b)\arctanh\bigg(\frac{\sqrt{2B}A}{(\sqrt{2}/b)A+(b/\sqrt{2})B}\bigg)
\end{equation}
where
\begin{equation}\label{solitarywave2}
A = 1+\sqrt{1-cU(\xi)^2},
\quad
B = b^2-1+\sqrt{1-cU(\xi)^2} .
\end{equation}
This algebraic equation determines $U(\xi)$ up to a sign,
and thus there are two smooth solitary wave solutions related by a reflection $U\leftrightarrow -U$, with each solution depending on the free parameter $0<b<1$. 

Finally, we will compare the smooth solitary wave solution $u=U(\xi)$
with the peakon travelling wave solution for equation \eqref{fg_singular_fam}. 

Peakon travelling waves have the form $u=a e^{-|x-ct|}$,
where $a$ is the amplitude and $c$ is the wave speed,
which satisfy an algebraic relation determined by a weak formulation of equation \eqref{fg_singular_fam}. 
Note that a peakon is a peaked travelling wave, 
but it is not a smooth solution and so it does not arise from the ODE \eqref{travelling_wave_ODE2}. 
Instead, applying the results in \Ref{AncRec} for the weak formulation of the general family \eqref{fg-fam} of multi-peakon equations, 
we find that $c = 1/a^2$ and hence 
\begin{equation}\label{peakon}
u = \pm (1/\sqrt{c}) e^{-|x-ct|} ,
\quad
c>0 . 
\end{equation}

This peakon travelling wave \eqref{peakon} 
and the smooth solitary wave \eqref{solitarywave1}--\eqref{solitarywave2}
have some similarities. 
Both move only in the positive $x$ direction, 
and both have either an up or down orientation. 
Their maximum amplitudes $|u|_{\max}$ are $1/\sqrt{c}$ for the peakon
and $b\sqrt{2-b^2}/\sqrt{c}$ for the smooth solitary wave,
both of which have the same dependence on the wave speed $c$. 
Note that the peak amplitude of the solitary wave is strictly less than that of the peakon,
since $b\sqrt{2-b^2}<1$ due to $b<1$. 

Remarkably, 
it seems that the singularity in equation \eqref{fg_singular_fam} 
affects just how the peak amplitude of travelling waves depends on the wave speed, 
with $|u|_{\max}\rightarrow \infty$ as $c\rightarrow 0$.

\section{Summary and concluding remarks}\label{concl}

In Theorems~\ref{thm:mom}, ~\ref{thm:H1}, and~\ref{thm:gradE}, 
we have presented classifications of families of nonlinear multi-peakon equations that possess conserved momentum \eqref{mom_integral}, 
conserved $H^1$ norm \eqref{H1_integral}, and conserved gradient energy \eqref{gradE_integral}.
These classifications include as special cases 
a weighted $H^2$ norm \eqref{wH2_integral} of $u$ 
and the $L^2$ norm \eqref{L2_integral} of $m$. 

As mentioned in section~\ref{intro}, 
these classifications reveal that there are two large families of equations 
for which, respectively, the $H^1$ norm of $u$ and the $L^2$ norm of $m$
are conserved. 
The first family is given by 
\begin{equation}
m_t + 2u_x h(u,u_x)m + u (h(u,u_x)m)_x=0
\end{equation}
where $h(u,u_x)$ is an arbitrary smooth function of $u$ and $u_x$;
the second family is given by 
\begin{equation}
m_t +\tfrac{1}{2}u_x h'(u)m + h(u)m_x =0
\end{equation}
where $h(u)$ is an arbitrary smooth function of $u$. 
In fact, conservation of the $H^1$ norm and the $L^2$ norm holds 
for slightly more general families, but the additional multi-peakon equations 
in those families contain singularities
when $u=0$, $u_x=0$, or $|u|=|u_x|$. 

We have considered three interesting applications of these classifications. 
Two of the applications look at special families of multi-peakon equations, 
one for which both the momentum and the $H^1$ norm are conserved,
and another for which the $H^1$ norm and the $L^2$ norm of $m$ are conserved. 
The first of these families is shown to have no smooth solitary wave solutions,
which significantly generalizes a result \cite{Len} 
known for the Camassa-Holm equation. 
In contrast, the second family is shown to possess 
smooth solitary wave solutions as well as peakon solutions. 
No other multi-peakon equation is known to have such a feature. 

There has been a lot of interest in recent years 
in studying integrable peakon equations. 
To-date, the known examples in the general class of multi-peakon equations \eqref{fg-fam} 
consist of the Camassa--Holm (CH) equation, the Degasperis--Procesi (DP) equation,
the Novikov (N) equation, and the modified Camassa--Holm (mCH) equation which is also known as the FORQ equation. 
A summary of which of these equations
possesses conservation of momentum, $H^1$ norm of $u$, $L^2$ norm of $m$, 
as well as conservation of a weighted $H^2$ norm of $u$, 
is presented in Table~\ref{table:integrable_peakon_eqns}. 
The latter result on weighted $H^2$ norms is not well known in the literature. 

\begin{table}[htb]
\centering
\caption{Integrable multi-peakon equations}
\label{table:integrable_peakon_eqns}
\begin{tabular}{c|cc|c|c|c|c}
\hline
Name & $f(u,u_x)$ & $g(u,u_x)$ & $||u||_{L^1}=||m||_{L^1}$ & $||u||_{H^1}$ & $||m||_{L^2}$ & $||u||_{H^2,\mu\neq2}$
\\
\hline
\hline
\tbox{Camassa-}{Holm}
&
$u_x$
& 
$u$
&
Y
&
Y
&
N
&
N
\\
\hline
\tbox{Degasperis--}{Procesi}
&
$2u_x$
& 
$u$
&
Y
&
N
&
N
&
N
\\
\hline
Novikov
&
$uu_x$
& 
$u^2$
&
N
&
Y
&
N
&
N
\\
\hline
\tbox{modified}{Camassa--Holm}
&
$0$
&
$u^2-u_x^2$
&
Y
&
Y
&
N
&
N
\\
\hline
\end{tabular}
\end{table}

Several recent papers 
\cite{MiMu,GraHim,AncSilFre,HimMan2016a,HimMan2016b,AncRecGanBru,AncRec}
have introduced nonlinear generalizations of 
the CH, DP, N, and mCH equations, involving an arbitrary nonlinearity power $p$.
Within the general class of multi-peakon equations \eqref{fg-fam}, 
the CH, DP, N, and mCH equations have been unified into a three-parameter family
given by \cite{HimMan2016a,HimMan2016b}
$f(u,u_x) = 3a(p-2) u^{p-3}u_x^3 + (b-p) u^{p-1}u_x$, 
$g(u,u_x) = u^{p} -3a u^{p-2} u_x^2$. 
This family can be slightly generalized by letting each term have an arbitrary coefficient:
\begin{equation}\label{CH-DP-N-mCH}
f(u,u_x) = \tilde a u^{p-3}u_x^3 + \tilde b u^{p-1}u_x,
\quad
g(u,u_x) = \tilde c u^{p-2} u_x^2 + \tilde d u^{p} . 
\end{equation}
The four-parameter family \eqref{CH-DP-N-mCH} contains 
a nonlinear CH-N $b$-family for $\tilde a=\tilde c=0$, $\tilde b = b$, $\tilde d=1$, 
\begin{equation}\label{CH-N-b-fam}
m_t + b u^{p-1}u_x m + u^{p} m_x =0  ,
\end{equation}
and a nonlinear mCH $b$-family for $\tilde a=\tilde b=0$, $\tilde c=1$, $\tilde d=b$, 
\begin{equation}\label{mCH-b-fam}
m_t + (u^{p-2}( u^2-b u_x^2)m)_x =0 .
\end{equation}

Although the generalized family \eqref{CH-DP-N-mCH} represents 
a unification of nonlinear generalizations of the CH, DP, N, and mCH equations, 
this family does not retain any of the integrability features of these equations. 

A nonlinear generalization unifying just the CH and mCH equations 
in a way that retains one of the Hamiltonian structures of both equations
is given by \cite{AncRec}
\begin{equation}\label{CH-mCH}
f(u,u_x) = a u_x(u^2-u_x^2)^{p/2},
\quad
g(u,u_x) = ( a u(u^2-u_x^2)^{p/2}  + b (u^2-u_x^2)^{(p+1)/2} . 
\end{equation}
This three-parameter family is a close analog of 
the gKdV equation which unifies the KdV and mKdV equations into a one-parameter family retaining one of the Hamiltonian structures shared by both equations. 
A one-parameter family of generalized CH (gCH) multi-peakon equations 
is obtained from this family \eqref{CH-mCH} 
by putting $k=2p-2$, $b=0$, $a=1$, which yields \cite{AncRec}
\begin{equation}\label{gCH}
m_t +  u_x (u^2-u_x^2)^{p-1} m + (u (u^2-u_x^2)^{p-1} m)_x = 0 . 
\end{equation}
Similarly, 
putting $k=2p-1$, $a=0$, $b=1$ in the three-parameter family \eqref{CH-mCH}
yields a one-parameter family of generalized mCH (gmCH) multi-peakon equations 
\cite{AncRec}
\begin{equation}\label{gmCH}
m_t + ((u^2-u_x^2)^p m)_x = 0 . 
\end{equation} 
Both of these one-parameter families \eqref{gCH} and \eqref{gmCH} 
have a Hamiltonian structure, unlike the generalized CH-DP-N-mCH family \eqref{CH-DP-N-mCH}.

We summarize in Table~\ref{table:peakon_eqns}
the conditions under which the non-Hamiltonian family \eqref{CH-DP-N-mCH}
and the Hamiltonian family \eqref{CH-mCH} 
possess conservation of momentum, $H^1$ norm of $u$, $L^2$ norm of $m$, 
as well as conservation of a weighted $H^2$ norm of $u$, 
by applying Theorems~\ref{thm:mom}, ~\ref{thm:H1}, and~\ref{thm:gradE}. 
One remark is that the Hamiltonian structure for the family \eqref{CH-mCH} 
corresponds to an energy conservation law that has a local density but a nonlocal flux, 
and hence it falls outside of the classifications considered 
in these classifications. 

\begin{table}[htb]
\centering
\caption{Families of multi-peakon equations}
\label{table:peakon_eqns}
\begin{tabular}{c|c|c|c|c}
\hline
Name & $||u||_{L^1}=||m||_{L^1}$ & $||u||_{H^1}$ & $||m||_{L^2}$ & $||u||_{H^2,\mu\neq2}$
\\
\hline
\hline
\tbox{g-CH-DP-}{N-mCH}
&
Y: $a=b=0$; 
&
Y: $a=d$, $b=c$;
&
Y: $a=d=0$, 
&
Y: $a=d=0$, 
\\
&
$2a=(1-p)b$, $p=1,3$
&
$a+b=c+d$, $p=2$
&
$2b=-pc$
&
$b=c$, $p=-2$
\\
\hline
\tbox{g-CH-mCH}{Hamiltonian}
&
Y
&
Y
&
N
&
N
\\
\hline
\end{tabular}
\end{table}

The nonlinear multi-peakon equations that possess 
conserved $H^1$ norm \eqref{H1_integral} of $u$, 
$L^2$ norm of $m$, and conserved gradient energy \eqref{gradE_integral} of $u$
comprise a wide family of equations which should be interesting to study further. 
In particular, 
an important goal would be to look at 
well-posedness and wave breaking criteria
in the Cauchy problem for these equations. 
Since controlling as many derivatives of $u$ as possible will be helpful, 
this motivates undertaking a search for additional conserved quantities, 
such as higher-derivative energies $E[u]=\int_\Omega m_x^2 + N(u,u_x,m)\, dx$. 
More generally, 
a classification of all local conservation laws of the form $T(u,u_x,m,m_x)$
would be of obvious interest.

\section*{Appendix: Euler operator equations}

The computations to find admitted local conservation laws
rely on the following result. 

\begin{lem}\label{euler_eqn}	
(i) A function 
\begin{equation}\label{k}
k=h_1(u,u_x)m+h_2(u,u_x)m^2+h_3(u,u_x)m^3
\end{equation}
satisfies $\hat E_{u}(k)=0$ 
iff 
\begin{equation}\label{h}
h_1 = u_x k_1(u^2-u_x^2) + k_0\frac{u}{u^2-u_x^2},
\quad
h_2=h_3=0 
\end{equation}
where $k_1$ is an arbitrary function of $u^2-u_x^2$, 
and $k_0$ is an arbitrary constant.
\\
(ii) Any function of the form \eqref{k}--\eqref{h} is a total $x$-derivative
\begin{equation}
k = 
D_x\Big( \tfrac{1}{2}K_1(u^2-u_x^2) + \tfrac{1}{2} k_0\ln\Big(\frac{u-u_x}{u+u_x}\Big)+k_0x \Big)
\end{equation}
where $K_1$ is a function of $u^2-u_x^2$ given by $K_1'=k_1$. 
\end{lem}

Proof of (i):
The equation $\hat E_{u}(k)=0$ 
splits with respect to the variables $m$, $m_x$, $m_{xx}$, 
which do not appear in $k$. 
This yields a linear system of determining equations for the functions $h_1$, $h_2$, $h_3$. 
After simplification, this system reduces to four equations:
\begin{gather}
h_2=0,
\quad
h_3=0,
\\
2h_1{}_u +u h_1{}_{u_x u_x}+ u_x h_1{}_{u u_x} =0,
\label{heq1}
\\
u_x^2 h_1{}_{u u} -u^2 h_1{}_{u_x u_x} -3u h_1{}_{u}  +u_x h_1{}_{u_x} -h_1 =0 . 
\label{heq2}
\end{gather}
The latter two equations are equivalent to the linear system that arises from splitting the equation $\hat E_{u}(h_1m)=0$. 

To solve equations \eqref{heq1} and \eqref{heq2}, 
we first combine equation \eqref{heq2} with $u$ times equation \eqref{heq1},
giving
\begin{equation}\label{heq3}
( u (h_1/u_x)_{u_x} + h_1{}_{u} )_{u} =0 .
\end{equation}
Similarly, we next combine this equation \eqref{heq3} times $-u_x/u$ with equation \eqref{heq2},
which yields
\begin{equation}\label{heq4}
( u_x^2(u (h_1/u_x)_{u_x} +h_1{}_{u}) )_{u_x} =0 .
\end{equation}
The resulting pair of equations \eqref{heq3}--\eqref{heq4}
can be directly integrated with respect to $u$ and $u_x$. 
This gives 
\begin{equation}
u (h_1/u_x)_{u_x} +h_1{}_{u}  = C/u_x^2,
\quad
C=\const
\end{equation}
which is a first-order linear PDE for $h_1$. 
The general solution is obtained by the method of characteristics. 
This completes the proof of (i).

Proof of (ii):
A direct evaluation of $D_x\big( \tfrac{1}{2}K_1(u^2-u_x^2) + \tfrac{1}{2}\ln((u-u_x)/(u+u_x)) +k_0x \big)$
yields $\big(u_x k_1(u^2-u_x^2) + k_0 u/(u^2-u_x^2)\big)m$ 
which is equal to $k$. 
This completes the proof of (ii).

\section*{Acknowledgements}
S.C.A. is supported by an NSERC research grant
and thanks the Department of Mathematics of University of C\'adiz
for additional support during the period in which this work was completed.


\begin{thebibliography}{999}

\bibitem{MiMu} 
Y. Mi and C. Mu, 
On the Cauchy problem for the modified Novikov equation with peakon solutions,
{\em J. Diff. Equ.} 254 (2013), 961--982.

\bibitem{GraHim}
G. Grayshan and A. Himonas,
Equations with peakon traveling wave solutions, 
{\em Adv. Dyn. Syst. and Appl.} 8 (2013) 217--232. 

\bibitem{AncSilFre} 
S.C. Anco, P.L. da Silva, and I.L. Freire, 
A family of wave-breaking equations generalizing the Camassa-Holm
and Novikov equations,
{\em J. Math. Phys.} 56 (2015), no. 9, 091506 (21pp). 

\bibitem{HimMan2016a}
A. Himonas and D. Mantzavinos,
An $ab$-family of equations with peakon travelling waves,
{\em Proc. Amer. Math. Soc.} 144 (2016), 3797--3811.

\bibitem{HimMan2016b}
A. Himonas and D. Mantzavinos,
The Cauchy problem for a 4-parameter family of equations with peakon travelling waves,
{\em Nonlin. Anal.} 133 (2016) 161--199.

\bibitem{AncRecGanBru} 
S.C. Anco, E. Recio, M.L. Gandarias, and M.S. Bruz\'on, 
A nonlinear generalization of the Camassa-Holm equation with peakon solutions,
in 
{\em Dynamical systems, Differential Equations and Applications}, 29--37, 
Proceedings of the 10th AIMS International Conference (Madrid, Spain), 2015. 

\bibitem{AncRec}
S.C. Anco, E. Recio, 
A general family of multi-peakon equations. 
arXiv: 1609.04354 math-ph 

\bibitem{Con-Esc-1998} 
A. Constantin and J. Escher, 
Wave breaking for nonlinear nonlocal shallow water equations,
{\em Acta Math.} 181 (1998), 229--243.

\bibitem{Esc-Liu-Yin}
J. Escher, Y. Liu, and Z. Yin, 
Global weak solutions and blow-up structure for the Degasperis-Procesi equation,
{\em J. Funct. Anal.} 241 (2006), 457--485. 

\bibitem{Liu-Yin}
Y. Liu and Z. Yin, 
Global existence and blow-up phenomena for the Degasperis-Procesi equation,
{\em Comm. Math. Phys.} 267 (2006), 801--820. 

\bibitem{Gui-Liu-Tian}
G. Gui, Y. Liu, and L. Tian, 
Global existence and blow-up phenomena for the peakon $b$-family of equations,
{\em Indiana Univ. Math. J.} 57 (2008), 1209--1233. 

\bibitem{Gui-Liu-Olv-Qu}
G. Gui, Y. Liu, P.J. Olver, and C. Qu, 
Wave-breaking and peakons for a modified Camassa-Holm equation, 
{\em Commun. Math. Phys.} 319 (2013), 731--759. 

\bibitem{Jia-Ni}
Z. Jiang and L. Ni, 
Blow-up phenomenon for the integrable Novikov equation, 
{\em J. Math. Anal. Appl.} 385 (2012), 551--558. 

\bibitem{AncBlu2002a}
S.C. Anco and G.W. Bluman, 
Direct construction method for conservation laws of partial differential equations. Part I: Examples of conservation law classifications,
{\em Euro. J. Appl. Math.} 13 (2002), 545--566. 

\bibitem{AncBlu2002b}
S.C. Anco and G.W. Bluman, 
Direct construction method for conservation laws of partial differential equations. Part II: General treatment, 
{\em Euro. J. Appl. Math.} 13 (2002), 567--585. 

\bibitem{Anc-review}
S.C. Anco,
Generalization of Noether's theorem in modern form to non-variational partial differential equations, 
in {\em Recent progress and Modern Challenges in Applied Mathematics, Modeling and Computational Science}, 119--182, 
Fields Institute Communications 79, 2017.

\bibitem{Olv-book}
P.J. Olver, 
{\em Applications of Lie Groups to Differential Equations},
Springer-Verlag, New York, 1993.

\bibitem{2ndbook}
G. Bluman, A. Cheviakov, S.C. Anco, 
{\em Applications of Symmetry Methods to Partial Differential Equations}, 
Springer Applied Mathematics Series 168, 
Springer: New York, 2010.

\bibitem{PopBih}
R.O. Popovych and A. Bihlo,
Inverse problem on conservation laws, 
arXiv: 1705.03547 (2017). 

\bibitem{ConEsc}
A. Constantin, J. Escher, 
Wave breaking for nonlinear nonlocal shallow water equations, 
{\em Acta. Math.} 181 (1998), 229--243. 

\bibitem{Len}
J. Lenells, 
Traveling wave solutions of the Camassa--Holm equation, 
{\em J. Differential Eqns.} 217(2), (2005), 393--430. 

\bibitem{PrzAnc}
M. Przedborski and S.C. Anco, 
Solitary waves and conservation laws for highly nonlinear wave equations modeling granular chains, 
{\em J. Math. Phys.}  58 (2017) 091502 (34 pages). 


\end{thebibliography}
\end{document}